\documentclass[pra,aps,twocolumn,showpacs,superscriptaddress]{revtex4}

\usepackage{graphicx}

\begin{document}

\title{Quantum cloning a pair of orthogonally polarized photons with linear optics}

\author{Jarom\'{\i}r Fiur\'{a}\v{s}ek}
\affiliation{Department of Optics, Palack\'{y} University, 
17. listopadu 50, 77200 Olomouc, Czech Republic}

\author{Nicolas J. Cerf\,}
\affiliation{QuIC, Ecole Polytechnique, CP 165, 
Universit\'{e} Libre de Bruxelles, 1050 Brussels, Belgium }

\begin{abstract}
A linear optical probabilistic scheme for the optimal cloning 
of a pair of orthogonally-polarized photons is devised, 
based on single- and two-photon interferences. It consists in
a partial symmetrization device, realized with a modified
unbalanced Mach-Zehnder interferometer, followed by two
independent Hong-Ou-Mandel interferometers. This scheme has
the advantage that it enables quantum cloning without the need 
for stimulated amplification in a nonlinear medium. It can also
be modified so to make an optical two-qubit partial SWAP gate,
thereby providing a potentially useful tool to linear optics 
quantum computing. 
\end{abstract}


\maketitle

\section{Introduction}

Perfect copying of unknown quantum states is forbidden due to the linearity of
quantum mechanics \cite{Wootters82,Dieks82}. The rapid development of quantum
information theory over the last decade has stimulated the investigation of 
optimal {\em approximate} quantum copying transformations,
which produce two or more copies of a state with maximum 
fidelity \cite{Cerf2006,Scarani2005}. The reason behind this interest
is twofold. First, the optimal ``quantum cloners'' provide
insight into the fundamental limits on the manipulation and distribution
of quantum information. Secondly, from a more practical point of view,
these cloners can be used as very efficient eavesdropping attacks
on quantum key distribution protocols. Of particular interest is the cloning
of single photons, which are ideal carriers
of quantum information as they can be distributed over long distances 
through optical fibers or free space.
In this context, the optimal universal copying of the polarization state
of single photons has been thoroughly investigated theoretically 
and successfully demonstrated experimentally by several groups. 
These experiments can be divided, roughly speaking, into two classes. 
The first strategy, suggested in Ref.~\cite{Simon2000}, consists of exploiting 
the amplification of light by means of stimulated parametric downconversion
in a non-linear crystal. It was demonstrated 
in Refs.~ \cite{Linares2002,DeMartini2004}. The second strategy is to
make the source photon interfere with an auxiliary photon prepared 
in a maximally mixed state on a beam splitter \cite{Ricci2004,Irvine2004,Khan2004,Masullo2004}. The bunching of photons 
then ensures the symmetrization of the total state of 
the photons, which is a way of effecting the optimal universal 
cloning transformation as shown in Ref.~\cite{Werner98}.

We thus observe that two fundamental quantum optical processes which are
quite unrelated, namely the amplification of light and the multiphoton interference, become interchangeable as far as quantum cloning is concerned. 
In this paper, we further exploit this interesting relationship
by designing an interferometric scheme for the optimal quantum copying 
of a pair of orthogonally polarized photons \cite{Fiurasek2002}. 
In the latter scenario, one assumes that the state to be cloned is formed 
by a pair of qubit states $|\psi\rangle|\psi_\perp\rangle$, where $\langle
\psi|\psi_\perp\rangle=0$ and $|\psi\rangle$ can be arbitrary. 
The optimal cloning operation which  produces $M$ copies of 
the state $|\psi\rangle$ from the state $|\psi\rangle|\psi_\perp\rangle$ 
with highest fidelity was derived in Ref.~\cite{Fiurasek2002}, where
it was shown that, surprisingly, the attained fidelity exceeds that of the
cloning of a pair $|\psi\rangle|\psi\rangle$ when $M>6$. It was also
proved that the optimal cloning of $|\psi\rangle|\psi_\perp\rangle$ could be
probabilistically accomplished with type-II non-degenerate parametric
down-conversion, similarly as for the cloning of a single photon. The trick,
however, is that the photon in polarization state $|\psi\rangle$ must be fed 
in the signal input port of the amplifier, while the photon 
in state $|\psi_\perp\rangle$ must be fed in the idler input port. 
With a certain probability, the amplifier produces $M$ copies of $|\psi\rangle$ in the output signal port,
with the value of $M$ being determined by measuring the number of
output idler photons. The distinct feature of this scheme with respect
to the original amplifier-based scheme, is that the  fidelity of the clones 
depends on the amplification gain, which has to be set to the
optimal value in order to recover the optimal cloning 
transformation \cite{Fiurasek2002}. 

In the present work, we show how to implement the universal cloning 
of the state $|\psi\rangle|\psi_\perp\rangle$ with passive linear optics
and auxiliary photons, circumventing the need for active non-linear media.
The term ``universal'' means that the transformation is independent
of the state $|\psi\rangle$, so that our scheme can be viewed as a way 
to effect the polarization-insensitive amplification of 
$|\psi\rangle|\psi_\perp\rangle$ without an optical amplifier.
In Section III, we will explain the working of our proposed scheme 
for the simplest yet nontrivial example of the generation of two copies; 
we will then extend it to arbitrary $M$.
A crucial part of our proposal happens to be the partial symmetrization 
of the two-photon polarization state. Therefore,
in Section II, we will first show how the partial symmetrization device
can be (probabilistically) accomplished by an interference of two photons
in a specifically designed Mach-Zehnder interferometer. Remarkably, 
a simple modification of this device also allows the realization
of a two-qubit optical partial-SWAP gate. Our novel scheme may thus find 
applications in various areas of quantum information processing 
with linear optics, beyond quantum cloning.

\section{Partial symmetrization device}

\begin{figure}[!t!]
\centerline{
\includegraphics[width=0.75\linewidth]{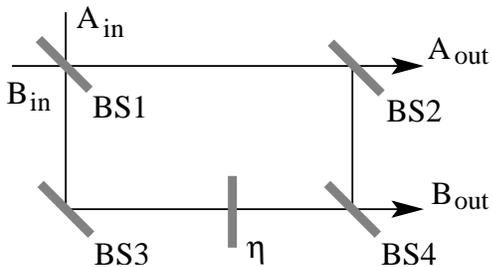}}
\caption{Partial symmetrization of the polarization state 
of two photons. The scheme consists of four balanced beam splitters BS$_j$
and one attenuator $\eta$.}
\end{figure}

Let us begin with the description of the partial symmetrization device. Let $\Pi_{+}$ and $\Pi_{-}$ denote projectors
onto the symmetric and anti-symmetric subspace of the 
polarization space of two photons. The non-unitary partial symmetrization operation is defined as
\begin{equation}
|\Psi\rangle_{AB} \mapsto (\Pi_{+}+ \eta \Pi_{-}) |\Psi\rangle_{AB},  
\label{partialsymmetrization}
\end{equation}
where $0 \leq \eta \leq 1$. The optical scheme which conditionally
implements this transformation is shown in Fig. 1.  
It is essentially a Mach-Zehnder interferometer made of 4 balanced beam
splitters BS1 -- BS4, with a variable attenuator with amplitude transmittance
$\eta$ placed in one of its arms. A single photon is injected into each input
port $A_\mathrm{in}$ and $B_\mathrm{in}$, and the partial symmetrization is
successful if a single photon is present in each output port $A_\mathrm{out}$ and $B_\mathrm{out}$. Note that these two output modes differ from 
those of a usual Mach-Zehnder interferometer since one output beam 
is obtained by tapping-off a part of the beam in the upper arm 
with the help of a balanced beam splitter BS2. 
The two input photons first interfere on the balanced beam splitter BS1,
which forms a Hong-Ou-Mandel interferometer. If the polarization state 
of the two photons is symmetric then bunching occurs and both photons 
end up in the same arm of the interferometer. With probability $1/2$, 
both photons follow the lower arm so that none 
can reach the output port $A_\mathrm{out}$, hence the device fails. 
However, with probability $1/2$, both photons are in the upper arm
and then, again with probability 1/2, one photon is reflected at BS2 
while the other is transmitted through BS2, ending in the output mode $A_\mathrm{out}$. Finally, the reflected photon can be again
reflected with probability $1/2$ at BS4, ending into
mode $B_\mathrm{out}$. Taking all these probabilities into account, 
we conclude that the symmetric part of the input polarization state
transforms according to 
\begin{equation}
\Pi_{+}|\Psi\rangle_{AB} \mapsto \frac{1}{2\sqrt{2}} \Pi_{+}|\Psi\rangle_{AB}.
\label{symmetricpart}
\end{equation}

Now, if the two-photon input state is antisymmetric, then the photons 
are in the maximally-entangled singlet state
$|\Psi^{-}\rangle_{AB}=\frac{1}{\sqrt{2}}(|V\rangle_A|H\rangle_B-|H\rangle_A|V\rangle_B)$. The photons never bunch and, after interference at BS1,
one photon is found in each arm of the interferometer so that 
the polarization state remains $|\Psi^{-}\rangle$. 
This well-known effect was exploited e.g. in quantum teleportation 
to carry out a partial Bell state analysis, that is to discriminate the
singlet from the other three Bell states \cite{Bouwmeester97}. In our scheme,
the photon in the upper arm is transmitted with probability $1/2$
through BS2 into mode $A_\mathrm{out}$. In parallel, with the
overall probability $\eta^2/4$, the photon in the lower arm 
is reflected at BS3, passes through the attenuator $\eta$ and BS4, and 
reaches mode $B_\mathrm{out}$. It follows that the antisymmetric part
of the input polarization state transforms according to
\begin{equation}
\Pi_{-}|\Psi\rangle_{AB} \mapsto \frac{\eta}{2\sqrt{2}} \Pi_{-} |\Psi\rangle_{AB}.
\label{antisymmetricpart}
\end{equation}
If the interferometer is balanced such that the photons traveling
through the upper and lower arms perfectly overlap at BS4, 
then the two above operations act coherently. Since an arbitrary 
input state can be decomposed into a symmetric and antisymmetric part,
$|\Psi\rangle_{AB}=\Pi_{+}|\Psi\rangle_{AB}+\Pi_{-}|\Psi\rangle_{AB}$,
the linearity of quantum mechanics implies that overall 
conditional transformation reads
\begin{equation}
|\Psi\rangle_{AB} \mapsto \frac{1}{2\sqrt{2}}(\Pi_{+}+\eta\Pi_{-}) |\Psi\rangle_{AB},
\label{Psitransformation}
\end{equation}
which is proportional to the desired partial symmetrization operation (\ref{partialsymmetrization}). The probability of success
generally depends on the input state and can be expressed as
\begin{equation}
P_{\mathrm{sym}}= \frac{1}{8}\langle \Psi| (\Pi_{+}+|\eta|^2 \Pi_{-}) |\Psi\rangle.
\label{Psucc}
\end{equation}
The fact that that the setup depicted in Fig.~1 effectively implements 
the partial symmetrization transformation (\ref{Psitransformation}) 
can also be verified by direct calculation in the Fock basis,
where the input two-photon state is written with the action of
bosonic creation operators onto the vacuum state, the creation operators 
of the input modes are replaced by appropriate linear combinations 
of the output creation operators, and finally only those terms are kept 
where a single photon is present in each output mode $A_\mathrm{out}$ 
and $B_\mathrm{out}$.

This suggested scheme is very versatile because the degree of symmetrization
can be controlled simply by changing the attenuation $\eta$ in one arm 
of the interferometer. The case with full attenuation corresponds
to the full symmetrization operation, while the case with no attenuation
simply effects the identity. Note that this partial
symmetrization scheme relies on a fine interplay between single- 
and two-photon interferences. This is in contrast with the partial
antisymmetrization device introduced in \cite{Filip2004}, which is defined 
in analogy with Eq.~(\ref{partialsymmetrization}) but interchanging 
the roles of $\Pi_{+}$ and $\Pi_{-}$. In that case, the device works
solely by a two-photon interference on an unbalanced beam splitter, 
whose reflectance determines the value of $\eta$ (for a 50:50 beam splitter,
we have full antisymmetrization). Our partial symmetrization device 
is thus experimentally more challenging,
but it opens interesting new perspectives as we shall see.

\section{Optimal quantum cloning without nonlinearities}

\subsection{Generation of 2 clones of $|\psi\rangle|\psi_\perp\rangle$}

Let us show how the above partial symmetrization device can be used
in order to optimally clone a pair of orthogonal 
qubits $|\psi\rangle|\psi_\perp\rangle$ with linear optics. 
In our scheme, qubits are encoded into polarization states of single photons.
We first consider the preparation of two clones, keeping the
case of $M$ clones for Sect. IIIC. Remember first how the 
polarization-insensitive cloning of single qubit $|\psi\rangle$ based
on linear optics works. The input photon in state $|\psi\rangle$
impinges on a balanced beam splitter BS1 where it interferes with 
another photon in a maximally-mixed polarization state (see Fig.~2a). 
The success of the cloning transformation is associated
with the bunching of the two photons at the beam splitter BS1, which
heralds the symmetrization operation. Thus, the cloning is 
witnessed by the detection of two photons in the output mode $A_{\mathrm{out}}$
of BS1. The state of the two clones of $|\psi\rangle$
is simply the polarization state of these two photons. Note that
the maximally-mixed state may be obtained by generating 
a two-photon polarization singlet state $|\Psi^-\rangle_{AB}$ (an EPR pair), 
sending one photon of this pair on BS1 (see Fig.~2a). Interestingly, 
if we postselect the other photon of the pair in the cases where 
we have detection of the two clones in $A_{\mathrm{out}}$, the polarization state
of this photon in mode $B$ then coincides with the so-called
anticlone, i.e., the approximate version of $|\psi_\perp\rangle$.

\begin{figure}[!t!]
\centerline{\includegraphics[width=0.75\linewidth]{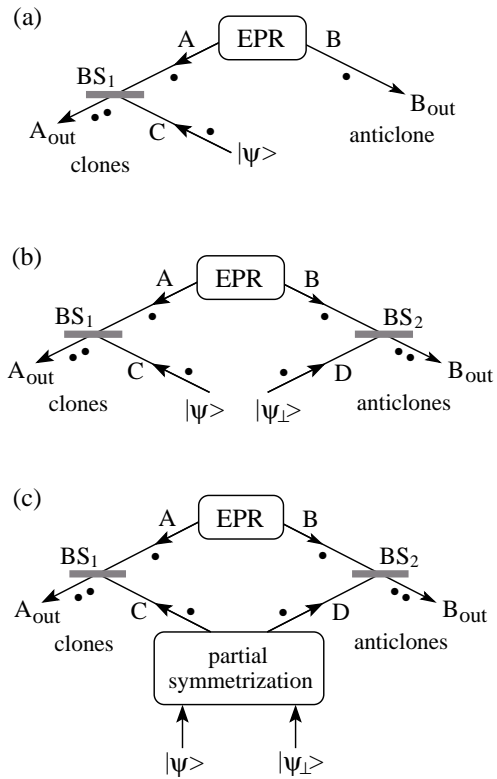}}
\caption{(a) Optimal cloning of the polarization state $|\psi\rangle$ of one photon
with linear optics. The EPR denotes an auxiliary two-photon polarization
singlet state, while BS1 is a balanced beam splitter.
(b) Suboptimal cloning of a pair of orthogonally polarized photons 
with linear optics. The second input is prepared in the orthogonal 
polarization state $|\psi_\perp\rangle$, and BS2 is another
balanced beam splitter. (c) Optimal cloning of the pair 
$|\psi\rangle |\psi_\perp\rangle$ using the previous cloning scheme
preceded by a partial symmetrization device. }
\end{figure}

Assume now that the second photon of the EPR pair (the one in mode $B$)
impinges on another balanced beam splitter BS2, where it interferes
with mode $D$ (see Fig.~2b). If the latter mode is in the vacuum state 
and we only keep the cases where the photon emerges in mode $B_{\mathrm{out}}$, the 
overall cloning transformation remains unchanged (except for a reduction
of the success probability by a factor of 1/2). However, this modification
is crucial because it introduces mode $D$ in the scheme, which plays
the same role as the input idler mode in the implementation of a cloner
based on the amplification of light. When using parametric stimulated 
downconversion in a nonlinear medium to effect cloning, one sends
a photon in state $|\psi\rangle$ in the signal mode of the amplifier, 
and leave the idler mode in the vacuum state. The two photons emerging
in the signal mode are the clones, while the photon in the output
idler mode encodes the anticlone. It was shown in Ref.~\cite{Fiurasek2002}
that the optimal cloning of a pair of orthogonally polarized photons 
$|\psi\rangle|\psi_\perp\rangle$
can be realized by injecting the photon in state $|\psi_\perp\rangle$ 
in the idler input mode instead of the vacuum. This improves
the fidelity of the cloning of the state $|\psi\rangle$ that is sent 
into the signal input mode.

This suggests that a linear optics version of the cloning of 
$|\psi\rangle|\psi_\perp\rangle$ may be achieved similarly by injecting
the photon in state $|\psi_\perp\rangle$ in mode $D$ instead of 
the vacuum (see Fig.~2b). The success of cloning is then heralded by
the detection of two photons in mode $A_{\mathrm{out}}$ (the two clones)
and two photons in mode $B_{\mathrm{out}}$ (the two anticlones).
The resulting probabilistic scheme approximates quite well the unphysical transformation $|\psi\rangle|\psi_\perp\rangle \mapsto |\psi\rangle^{\otimes 2}
|\psi_\perp\rangle^{\otimes 2}$. However,  as we will see below, 
an additional element is needed in order to achieve the optimal cloning
transformation, namely the partial symmetrization device described 
in Sect.~II.

The final scheme, which realizes the optimal cloning,
is illustrated in Fig.~2c. The two-photon input state to be cloned 
$|\psi\rangle|\psi_\perp\rangle$ is first partially symmetrized 
using the setup of Fig.~1. The resulting state in modes $C$ and $D$ reads
\begin{eqnarray}
|\Psi_\mathrm{in}\rangle_{CD}&=&(\Pi_{+}+\eta
\Pi_{-})|\psi\rangle_C|\psi_{\perp}\rangle_D \nonumber \\
&=&\frac{1+\eta}{2}|\psi\rangle_C|\psi_\perp\rangle_D+
\frac{1-\eta}{2}|\psi_\perp\rangle_C|\psi\rangle_D,
\label{PsiinCD}
\end{eqnarray}
where we made use of the relations
$\Pi_{+}|\psi\rangle|\psi_\perp\rangle=(|\psi\rangle|\psi_\perp\rangle+|\psi_\perp\rangle|\psi\rangle)/2$
and
$\Pi_{-}|\psi\rangle|\psi_\perp\rangle=(|\psi\rangle|\psi_\perp\rangle-|\psi_\perp\rangle|\psi\rangle)/2$.
As in the previous scheme, this state then interferes 
on the two balanced beam splitters BS1 and BS2 with the auxiliary 
two-photon singlet state $|\Psi^{-}\rangle_{AB}$. 

With the help of creation operators,
we can express the states impinging on BS1 and BS2 as
\begin{equation}
|\Psi^{-}\rangle_{AB}=
\frac{1}{\sqrt{2}}(a^\dagger_{\psi}b_{\psi_\perp}^\dagger-a^\dagger_{\psi_\perp}b_{\psi}^\dagger)|0\rangle.
\label{Psiminuscreation}
\end{equation}
and 
\begin{equation}
|\Psi_\mathrm{in}\rangle_{CD}= \frac{1}{2}[(1+\eta)c_{\psi}^\dagger d_{\psi_\perp}^\dagger+
(1-\eta)c_{\psi_\perp}^\dagger d_{\psi}^\dagger] |0\rangle,
\label{Psiincreation}
\end{equation}
where the subscript $\psi$ ($\psi_\perp$) indicates creation operator 
for the polarization mode $\psi$ ($\psi_\perp$).
The cloning operation is probabilistic and succeeds if two photons are present in output mode $A_\mathrm{out}$ and two photons are present in
$B_\mathrm{out}$. Indeed, in that case the state of photons in modes A and C 
(B and D) is thus symmetrized by the interference on BS1 (BS2). 
The output state can be determined from the input state $|\Psi^{-}\rangle_{AB}|\Psi_\mathrm{in}\rangle_{CD}$
by replacing the input creation operators with the appropriate linear
combinations of output creation operators, namely 
\begin{equation}
\begin{array}{lcl}
a_{\psi}^{\dagger} \mapsto \frac{1}{\sqrt{2}}(a_\psi^\dagger+c_\psi^\dagger), & \quad &
b_{\psi}^{\dagger} \mapsto \frac{1}{\sqrt{2}}(b_\psi^\dagger+d_\psi^\dagger), \\
c_{\psi}^{\dagger} \mapsto \frac{1}{\sqrt{2}}(a_\psi^\dagger-c_\psi^\dagger),  & &
d_{\psi}^{\dagger} \mapsto \frac{1}{\sqrt{2}}(b_\psi^\dagger-d_\psi^\dagger), 
\end{array}
\label{abcdtransformation}
\end{equation}
while similar relations hold for the creation operators 
for the $\psi_\perp$ polarization modes. By keeping only the terms
of zeroth order in $c_{\psi}^{\dagger}$, $c_{\psi_\perp}^{\dagger}$,
$d_{\psi}^{\dagger}$, and $d_{\psi_\perp}^{\dagger}$, 
we obtain the following expression for the (unnormalized) output state 
conditional on having two photons in mode $A_\mathrm{out}$ and two photons in mode $B_\mathrm{out}$,
\begin{equation}
|\Psi_{2,\mathrm{out}}\rangle \propto
(a_\psi^\dagger b_{\psi_\perp}^\dagger -a_{\psi_\perp}^\dagger b_\psi^\dagger)
(a_\psi^\dagger b_{\psi_\perp}^\dagger + q \; a_{\psi_\perp}^\dagger b_\psi^\dagger)
|0\rangle,
\label{Psi2out}
\end{equation}
where $q\equiv (1-\eta)/(1+\eta)$ has been introduced 
for the sake of simplicity. Omitting the subscripts $\psi$ and $\psi^\dagger$,
we define $|j,k\rangle$ as the double Fock state containing $j$ photons
in polarization state $|\psi\rangle$ and $k$ photons 
in polarization state $|\psi_\perp\rangle$.
In particular, $|\psi\rangle_A|\psi_\perp\rangle_B$ becomes  
$|1,0\rangle_A |0,1\rangle_B$.
Then, the output state (\ref{Psi2out}) can be expressed as a linear
combination of double Fock states,
\begin{eqnarray}
|\Psi_{2,\mathrm{out}}\rangle & \propto & 2 \; |2,0\rangle_A  |0,2\rangle_B
+(q-1)\; |1,1\rangle_A |1,1\rangle_B \nonumber \\
& & -2q \; |0,2\rangle_A |2,0\rangle_B.
\label{Psi2outexpanded}
\end{eqnarray}
The first term in the right-hand side of Eq.~(\ref{Psi2outexpanded}) 
corresponds to perfect cloning, 
with two photons in mode $A_\mathrm{out}$ emerging in
the polarization state $|\psi\rangle$ and
two photons in mode $B_\mathrm{out}$ being in
state $|\psi_\perp\rangle$. In the second term, one photon has the right
polarization and one photon has the wrong polarization in both 
$A_\mathrm{out}$ and $B_\mathrm{out}$, thus contributing to $1/2$ in the cloning
fidelity. The third term does not contribute to the cloning fidelity 
since the two photons have a polarization state that is orthogonal to the expected one in each output mode. Thus,
the fidelity of single clones in mode $A_\mathrm{out}$ (or single anticlones
in mode $B_\mathrm{out}$) can be evaluated as 
\begin{equation}
F_\perp(2,q)=\frac{1\times 2^2+\frac{1}{2}\times (q-1)^2}{2^2+(q-1)^2+(2q)^2}=
\frac{q^2-2q+9}{2(5q^2-2q+5)}.
\label{F2q}
\end{equation}
Note that in deriving formula (\ref{F2q}), we have taken into account that 
the state (\ref{Psi2outexpanded}) is not normalized.

Coming back to the simplified cloning scheme of Fig.~2b, we see
that removing the partial symmetrization device is equivalent to
taking $\eta=1$ (or $q=0$) in the scheme of Fig.~2c. 
In that case, the first term of Eq.~(\ref{Psi2outexpanded}) weighs
$2^2$ and the second term $1^2$, while the third term vanishes.
The resulting single-clone fidelity is 9/10. 
We notice that by choosing a small value of $q>0$,
the weight of the second term decreases linearly with $q$ while that of
the third term increases only quadratically with $q$. Since these 
two terms correspond to cloning noise, it is clear that increasing $q$ 
is advantageous as it decreases the overall noise up to some extent.
The optimal $q$ that maximizes $F_\perp(2,q)$ (or minimizes the noise)
can be determined by solving the equation  
$\frac{\partial F_\perp(2,q)}{\partial q}=0$, which yields
\begin{equation}
q_{2,\mathrm{opt}}=5-2\sqrt{6}.
\label{q2opt}
\end{equation}
By inserting $q_{2,\mathrm{opt}}$ back into Eq. (\ref{F2q}), 
we obtain 
\begin{equation}
F_\perp(2)=\frac{1}{2}\left(1+\sqrt{\frac{2}{3}}\right) \simeq 0.908
\end{equation}
which is the maximum achievable fidelity \cite{Fiurasek2002}. 
Our scheme thus optimally clones the pair of orthogonal photonic
qubits, as advertised. Note that the achieved fidelity is indeed
slightly larger than that of the simplified scheme which is not preceded
by the partial symmetrization of $|\psi\rangle|\psi_\perp\rangle$, namely 9/10.
Note also that the optimal attenuation corresponding 
to the value of $q$ of Eq.~(\ref{q2opt}) reads
$\eta_{2,\mathrm{opt}}=\sqrt{2/3}$. The total probability of success 
of the cloning scheme is given by 
$P_\mathrm{tot}=P_\mathrm{sym}P_\mathrm{EPR}$, where $P_\mathrm{sym}=5/48$ 
is the probability of
successful partial symmetrization and $P_\mathrm{EPR}=3/20$ is the probability of success of the scheme in Fig. 2b. Hence, we have $P_\mathrm{tot}=1/64$.

\subsection{Experimental feasibility of the proposed setup}

Let us briefly discuss how the setup of Fig 2c may be demonstrated experimentally. 
First note that although the partial symmetrization device operates on a
coincidence basis, this does not negatively affect the rest of the 
scheme shown in Fig. 2c. This is because after partial symmetrization,
the two emerging photons are not recombined at any further beam splitter,
but participate each in another separate two-photon interference. 
Thus, if we know that a two-photon singlet
state was injected into input modes A and B, then detection of two photons in
modes $A_\mathrm{out}$ and $B_\mathrm{out}$ confirms the successful partial symmetrization.
To prepare the two-photon input state one could utilize for example
two single photons, each one being prepared conditionally from a photon pair generated via spontaneous parametric downconversion by triggering on 
detections of the idler photons. 
The auxiliary singlet state in modes A and B could be prepared 
in the same way. The whole experiment would then involve a six-photon coincidence. Even if very challenging, recent experiments with three
photon pairs and six-fold coincidence measurements were reported
\cite{Zhang2006,Lu2006}, so our proposed scheme is within
the reach of current technology. 

The partial symmetrization device requires interferometric
stability. Recently, interference of two photons in bulk Mach-Zehnder interferometer was demonstrated experimentally and explored for the optimal universal asymmetric cloning of single photons \cite{Zhao2005}. Moreover, the combination of single- and two-photon interferences has recently been
utilized in fiber-based experiments, where very high visibility was
achieved \cite{Bartuskova2006a,Bartuskova2006b}. This suggests that
our cloning scheme is experimentally realizable.

An interesting feature of our scheme is that the nonlinearity
which is inherent to the cloner when realized with stimulated amplification
is hidden here in the prior preparation of the EPR pair. This can be viewed
as a nonlinear resource which is prepared ``off-line'', and used
only later on when needed, which is reminiscent to the idea behind
linear-optics quantum computing \cite{Knill2001,Kok2007}.

\subsection{Generation of $M$ clones of $|\psi\rangle|\psi_\perp\rangle$}

Let us finally show how our scheme can be extended to more than two clones.
The optimal copying operation which produces $M$ copies of $|\psi\rangle$
from the state $|\psi\rangle|\psi_\perp\rangle$  can be written in a covariant form as
follows \cite{Fiurasek2002},
\begin{equation}
|\psi\rangle_A|\psi_\perp\rangle_B \mapsto \sum_{j=0}^M
\alpha_{j,M}|(M-j)\psi,j\psi_\perp\rangle_A|j\psi,(M-j)\psi_\perp\rangle_B,
\label{cloningtransformation}
\end{equation}
where the coefficients $\alpha_{j,M}$ can be expressed as
\begin{equation}
\alpha_{j,M}=\frac{(-1)^j}{\sqrt{2(M+1)}}\left[1+\frac{\sqrt{3}(M-2j)}{\sqrt{M(M+2)}}\right]
\label{alphajM}
\end{equation}
and $|j\psi,(M-j)\psi_\perp\rangle$ denotes a totally symmetric polarization 
state of $M$ photons in a single spatiotemporal mode, 
with $j$ photons in polarization state $|\psi\rangle$ 
and $M-j$ photons in state $|\psi_\perp\rangle$.
The transformation (\ref{cloningtransformation}) achieves the fidelity
\begin{equation}
F_\perp(M)= \frac{1}{2}\left( 1+ \sqrt{\frac{M+2}{3M}}\right).
\label{Fperp}
\end{equation}

A physical insight into the structure of the unitary transformation
(\ref{cloningtransformation}) is obtained by
considering cloning via stimulated amplification. With a suitable choice of
the crystal geometry, the type-II parametric down-conversion is governed by 
the Hamiltonian 
\begin{equation}
H=ig(a_V^\dagger b_H^\dagger-a_H^\dagger b_V^\dagger)+\mathrm{h.c.},
\label{Hamiltonian}
\end{equation}
where $a_V^\dagger$ $(a_H^\dagger)$ and $b_V^\dagger$ $(b_H^\dagger)$ 
denote creation operators of the vertically (horizontally) polarized modes
of the signal and idler beam, respectively, and $g$ is the effective amplification gain, proportional to the amplitude of the pump beam, 
the second-order susceptibility of the crystal, and
the crystal length. Since $H$ is invariant under simultaneous identical
local polarization rotation of both the signal and idler beams, 
$(U\otimes U) H (U^\dagger \otimes U^\dagger)=H$, we can write 
\begin{equation}
H=ig\; (a_\psi^\dagger b_{\psi_\perp}^\dagger-a_{\psi_\perp}^\dagger
b_{\psi}^\dagger)+\mathrm{h.c.}
\end{equation}
The output state of the amplifier reads
$|\Psi_{M,\mathrm{out}}\rangle=e^{-iH}|\psi\rangle_A|\psi_\perp\rangle_B$,
where $A$ stands for the signal mode and $B$ for the idler mode.
The unitary transformation can be conveniently written in a factorized form,
\begin{equation}
e^{-iH}=e^{\lambda X_\psi} (1-\lambda^2 )^{n_\mathrm{tot}/2+1}e^{-\lambda
X_\psi^\dagger},
\label{Unitary}
\end{equation}
where $\lambda=\tanh(g)$, $X_\psi=a_\psi^\dagger b_{\psi_\perp}^\dagger
-a_{\psi_\perp}^\dagger b_\psi^\dagger$ and $n_\mathrm{tot}$
denotes the total photon-number operator in all four modes, 
$n_\mathrm{tot}=a_\psi^\dagger a_\psi+ a_{\psi_\perp}^\dagger a_{\psi_\perp} 
+b_\psi^\dagger b_\psi +b_{\psi_\perp}^\dagger b_{\psi_\perp}.$
Using Eq.~(\ref{Unitary}), it can be shown by 
a straightforward calculation that the output state corresponding 
to having $M$ photons in each mode (signal and idler) reads 
\begin{equation}
|\Psi_{M,\mathrm{out}}\rangle \propto
X_\psi^{M-1} 
(a_\psi^\dagger b_{\psi_\perp}^\dagger + q \; a_{\psi_\perp}^\dagger b_\psi^\dagger) |0\rangle,
\label{PsiMout}
\end{equation}
where the coefficient $q$ depends on the gain $g$. For each value of $M$,
there is an optimal $q$ which maximizes the cloning fidelity,
$q_{M,\mathrm{opt}}=(\sqrt{3M}-\sqrt{M+2})/(\sqrt{3M}+\sqrt{M+2})$.

Finally, comparing Eq.~(\ref{PsiMout}) with Eq.~(\ref{Psi2out})
suggests that the state $|\Psi_{M,\mathrm{out}}\rangle$ can be similarly
prepared using the scheme shown in Fig.~2c, where the input state 
is partially symmetrized with factor $\eta=(1-q)/(1+q)$ while the EPR state
is replaced by the $2(M-1)$-photon state produced by spontaneous type-II parametric down-conversion, 
$|\Psi_\mathrm{EPR}\rangle \propto X_\psi^{M-1}|0\rangle$. 
Note that this state, corresponding to $(M-1)$ photon pairs,
can be generated ``off-line'' in a nonlinear crystal with low gain $g$, 
as in the case of $M=2$. In contrast, the implementation of this
cloner with stimulated amplification requires a gain $g$ which grows
with $M$. Interestingly, we thus conclude that 
the above linear optical scheme enables us to simulate the 
polarization-insensitive amplification of 
$|\psi\rangle|\psi_\perp\rangle$ with arbitrarily 
high gain using only a ``off-line'' low-gain nonlinear process.

\section{Conclusions}

It was shown that the optimal quantum cloning of a pair of photons 
with orthogonal  polarizations $|\psi\rangle|\psi_\perp\rangle$
can be experimentally realized by using a linear optical scheme, 
avoiding the traditional use of stimulated 
parametric downconversion in a nonlinear crystal. An appropriate
combination of one- and two-photon interferences makes it possible
to effect, with some success probability, the transformation resulting
into $M$ clones and $M$ anticlones 
from the pair $|\psi\rangle|\psi_\perp\rangle$.
This transformation consists in the sequence of a partial
symmetrization device, acting on the input state $|\psi\rangle|\psi_\perp\rangle$, followed by two parallel
Hong-Ou-Mandel interferometers (with the bunching effect being
used as a means to symmetrize the state).

The partial symmetrization device is another probabilistic
interferometric scheme, with an attenuator $\eta$ placed in one
of the interferometer arms in order to tune the symmetrization parameter.
Moreover, if this attenuator is replaced by a phase shifter, 
it appears that the symmetrization scheme instead effects
the unitary partial SWAP gate 
(with probability $1/8$). Formally, we can substitute $\eta=e^{i\phi}$ 
in Eq.~(\ref{partialsymmetrization}), and get the unitary transformation
\begin{equation}
U(\phi)=\Pi_{+}+e^{i\phi}\Pi_{-}.
\label{Uphi}
\end{equation}
For $\phi=\pi$, we recover the SWAP gate, which interchanges 
the states of the two photons. Of particular importance is 
the square-root SWAP gate, which is achieved by choosing 
$\phi=\pi/2$. This gate, together with arbitrary single-qubit
polarization rotations, is sufficient for universal quantum computing.  
Thus, we have found a whole new class of two-qubit optical quantum gates
to be inserted in the toolbox of available linear-optics quantum gates
\cite{Knill2001,Kok2007}.

\acknowledgments

JF acknowledges support from the Ministry of Education 
of the Czech Republic under projects ``Centre of Modern Optics'' 
(LC06007)  and ``Measurement and Information in Optics'' (MSM6198959213). 
NJC acknowledges financial support from the European Union under the
integrated project {\tt QAP}, from Brussels region government 
under project {\tt CRYPTASC}, and from the IUAP programme 
of the Belgian federal government under project {\tt photonics@be}.

\end{document}